\documentclass[11pt,a4paper]{article}

\pdfoutput=1

\usepackage{jheppub}

\usepackage{graphicx}
\usepackage{ulem}
\usepackage{color}

\def\beq{\begin{equation}}
\def\eeq{\end{equation}}
\def\bea{\begin{eqnarray}}
\def\eea{\end{eqnarray}}
\def\nn{\nonumber}

\def\roughly#1{\mathrel{\raise.3ex\hbox
{$#1$\kern-.75em\lower1ex\hbox{$\sim$}}}}
\def\lsim{\roughly<}
\def\gsim{\roughly>}
\def\sla#1{\raise.15ex\hbox{$/$}\kern-.57em #1}
\def\bra#1{\left\langle #1\right|}
\def\ket#1{\left| #1\right\rangle}

\def\bd{B_d^0}
\def\bs{B_s^0}
\def\bsbar{{\bar B}^0_s}

\def\btos{{\bar b} \to {\bar s}}
\def\ANPq{{\cal A}^q}

\def\ApNPqph{{\cal A}^{\prime,q} e^{i \Phi'_q}}
\def\ApNPCqph{{\cal A}^{\prime {C}, q} e^{i \Phi_q^{\prime C}}}
\def\ApNPCuph{{\cal A}^{\prime {C}, u} e^{i \Phi_u^{\prime C}}}
\def\ApNPCdph{{\cal A}^{\prime {C}, d} e^{i \Phi_d^{\prime C}}}
\def\pewcp{P_{EW}^{\prime C}}
\def\pewp{P'_{EW}}
\def\pewcpnp{P_{EW, NP}^{\prime C}}
\def\pewpnp{P'_{EW, NP}}
\def\ApNPuph{{\cal A}^{\prime,u} e^{i \Phi'_u}}
\def\ApNPdph{{\cal A}^{\prime,d} e^{i \Phi'_d}}
\def\ApNPcomb{{\cal A}^{\prime, comb} e^{i \Phi'}}
\def\btopik{B \to \pi K}
\def\bsmumu{b \to s \mu^+ \mu^-}

\def\bsqq{b \to s {\bar q} q}
\def\bsdd{b \to s {\bar d} d}
\def\bsuu{b \to s {\bar u} u}

\begin{document}

\title{\boldmath The $\btopik$ Puzzle Revisited}

\author[a]{Nicolas Boisvert Beaudry,}
\author[b,c]{Alakabha Datta,}
\author[a]{David London,}
\author[b,d]{Ahmed Rashed,} 
\author[a]{and Jean-Samuel Roux}
\affiliation[a]{Physique des Particules, Universit\'e de Montr\'eal, \\
C.P. 6128, succ. centre-ville, Montr\'eal, QC, Canada H3C 3J7,}
\affiliation[b]{Department of Physics and Astronomy, \\
108 Lewis Hall, University of Mississippi, Oxford, MS 38677-1848, USA, }
\affiliation[c]{Department of Physics and Astronomy, \\
 2505 Correa Rd, University of Hawaii, Honolulu, Hi 96826, USA, }
\affiliation[d]{Department of Physics, Faculty of Science, Ain Shams University, Cairo, 11566, Egypt,}
\emailAdd{nicolas.boisvert-beaudry@umontreal.ca}
\emailAdd{datta@phy.olemiss.edu}
\emailAdd{london@lps.umontreal.ca}
\emailAdd{amrashed@go.olemiss.edu}
\emailAdd{jean-samuel.roux@umontreal.ca}

\abstract{For a number of years, there has been a certain
  inconsistency among the measurements of the branching ratios and CP
  asymmetries of the four $\btopik$ decays ($B^+ \to \pi^+ K^0$, $B^+
  \to \pi^0 K^+$, $\bd \to \pi^- K^+$, $\bd \to \pi^0 K^0$). In this
  paper, we re-examine this $\btopik$ puzzle. We find that the key
  unknown parameter is $|C'/T'|$, the ratio of color-suppressed and
  color-allowed tree amplitudes. If this ratio is large, $|C'/T'| =
  0.5$, the SM can explain the data. But if it is small, $|C'/T'| =
  0.2$, the SM cannot explain the $\btopik$ puzzle -- new physics (NP)
  is needed. The two types of NP that can contribute to $\btopik$ at
  tree level are $Z'$ bosons and diquarks. $Z'$ models can explain the
  puzzle if the $Z'$ couples to right-handed $u{\bar u}$ and/or
  $d{\bar d}$, with $g_R^{dd} \ne g_R^{uu}$. Interestingly, half of
  the many $Z'$ models proposed to explain the present anomalies in
  $\bsmumu$ decays have the required $Z'$ couplings to $u{\bar u}$
  and/or $d{\bar d}$. Such models could potentially explain both the
  $\bsmumu$ anomalies and the $\btopik$ puzzle. The addition of a
  color sextet diquark that couples to $ud$ can also explain the
  puzzle.}


\arxivnumber{1709.07142}

\preprint{
{\flushright
UdeM-GPP-TH-17-258 \\
UMISS-HEP-2017-04 \\
}}

\maketitle

\section{Introduction}
\label{Sec:Intro}

Over the past 15 years there has been a great deal of interest in
$\btopik$ decays. There are four such processes: $B^+ \to \pi^+ K^0$
(designated as $+0$ below), $B^+ \to \pi^0 K^+$ ($0+$), $\bd \to \pi^-
K^+$ ($-+$) and $\bd \to \pi^0 K^0$ ($00$). Their amplitudes are not
independent, but obey a quadrilateral isospin relation:
\beq
\sqrt{2} A^{00} + A^{-+} = \sqrt{2} A^{0+} + A^{+0} ~.
\eeq
Using these decays, nine observables have been measured: the four
branching ratios, the four direct CP asymmetries $A_{CP}$, and the
mixing-induced indirect CP asymmetry $S_{CP}$ in $\bd\to
\pi^0K^0$. Shortly after these measurements were first made (in the
early 2000s), it was noted that there was an inconsistency among
them. This was referred to as the ``$\btopik$ puzzle'' \cite{BFRS1,
  BFRS2, BFRS3}.

In Ref.~\cite{Baek:2004rp}, it was pointed out that, by performing a
full fit to the data, one can quantify the discrepancy with the
standard model (SM) implied by the $\btopik$ puzzle.
Ref.~\cite{Baek:2004rp} (2004) contains the results of the first such
fit, an update was done in Ref.~\cite{Baek:2007yy} (2007), and the
last fit was performed in Ref.~\cite{Baek:2009pa} (2009).  This latter
study finds that ``if one adds a constraint on the weak phase $\gamma$
coming from independent measurements -- the SM fit -- one finds that
the fit is poor.  On the other hand, it is not terrible.  If one is
willing to accept some deficiencies in the fit, it can be argued that
the SM can explain the $\btopik$ data.'' In other words, one cannot
say that the $\btopik$ puzzle suggests the presence of new physics
(NP). A more correct statement would be that the measurements of
$\btopik$ decays {\it allow} for NP.

In the present paper, we update the fit using the latest experimental
results. As the data has not changed enormously since 2007, we do not
expect to find a different conclusion. However, the first purpose of
our study is to be more precise in this conclusion. In what regions of
parameter space does the SM yield a reasonable explanation of the
$\btopik$ data?  And in what regions is the SM fit to the data
terrible, so that NP is definitely required?

There are two types of NP that can contribute at tree level to
$\btopik$. One is a $Z'$ boson that has a flavor-changing coupling to
${\bar s}b$ and also couples to ${\bar u}u$ and/or ${\bar d}d$. The
other is a diquark that has $db$ and $ds$ couplings or $ub$ and $us$
couplings. The second goal of our work is to determine what precise
couplings the $Z'$ or diquark must have in order to improve the fit
with the $\btopik$ data.

Finally, we make one other observation. Over the past few years,
several measurements have been made that disagree with the predictions
of the SM. These include (i) $R_K$ (LHCb \cite{RKexpt}) and (ii)
$R_{K^*}$ (LHCb \cite{RK*expt}), where $R_{K,K^*} \equiv {\cal
  B}(B^{+,0} \to K^{+,*0} \mu^+ \mu^-)/{\cal B}(B^{+,0} \to K^{+,*0}
e^+ e^-)$, (iii) the angular distribution of $B \to K^* \mu^+\mu^-$
(LHCb \cite{BK*mumuLHCb1, BK*mumuLHCb2}, Belle \cite{BK*mumuBelle},
ATLAS \cite{BK*mumuATLAS} and CMS \cite{BK*mumuCMS}), and (iv) the
branching fraction and angular distribution of $\bs \to \phi \mu^+
\mu^-$ (LHCb \cite{BsphimumuLHCb1, BsphimumuLHCb2}). Recent analyses
of these discrepancies \cite{Capdevila:2017bsm, Altmannshofer:2017yso,
  DAmico:2017mtc, Hiller:2017bzc, Geng:2017svp, Ciuchini:2017mik,
  Celis:2017doq, Ghosh:2017ber, Alok:2017sui, Wang:2017mrd,
  Datta:2017ezo} combine constraints from all measurements and come to
the following conclusions: (i) there is indeed a significant
disagreement with the SM, somewhere in the range of 4-6$\sigma$, and
(ii) the most probable explanation is that the NP primarily affects
$\bsmumu$ transitions.  Arguably the simplest NP explanation is that
its contribution to $\bsmumu$ comes from the tree-level exchange of a
$Z'$ boson that has a flavor-changing coupling to ${\bar s} b$, and
also couples to $\mu^+ \mu^-$. Many models of this type have been
proposed to explain the data. In some of these models, the $Z'$ also
has couplings to ${\bar u} u$ and/or ${\bar d} d$. In this case, it
will also contribute at tree level to $\btopik$
decays\footnote{Another class of models involves the tree-level
  exchange of a leptoquark (LQ). However, LQs cannot contribute to
  $\btopik$ at tree level.}  and could potentially furnish an
explanation of the $\btopik$ puzzle. If so, in such models there would
be a connection between the anomalies in $\bsmumu$ and $\btopik$
decays, which is quite intriguing.

In Sec.~2 we review the $\btopik$ puzzle and update its status. In
Sec.~3 we examine whether the SM can explain the puzzle. We find that
it can if $|C'/T'| = 0.5$, but not if $|C'/T'| = 0.2$ (both values are
allowed theoretically). For the case of $|C'/T'| = 0.2$, in Sec.~4 we
examine the puzzle in the presence of NP.  We find that the $\btopik$
puzzle can be explained if the NP is a $Z'$ boson or a diquark.  We
make the connection with $Z'$ models that explain the $\bsmumu$
anomalies. We conclude in Sec.~5.

\section{\boldmath The $\btopik$ Puzzle}

We begin by reviewing the $\btopik$ puzzle. Within the diagrammatic
approach \cite{GHLR1, GHLR2}, $B$-decay amplitudes are expressed in
terms of six diagrams\footnote{We neglect the annihilation, exchange
  and penguin-annihilation diagrams, which are expected to be very
  small in the SM.}: the color-favored and color-suppressed tree
amplitudes $T'$ and $C'$, the gluonic penguin amplitudes $P'_{tc}$ and
$P'_{uc}$, and the color-favored and color-suppressed electroweak
penguin amplitudes $\pewp$ and $\pewcp$.  (The primes on the
amplitudes indicate $\btos$ transitions.) The $\btopik$ decay
amplitudes are given in terms of diagrams by
\bea
\label{fulldiagrams}
A^{+0} &=& -P'_{tc} + P'_{uc} e^{i\gamma} -\frac13
\pewcp ~, \nn\\
\sqrt{2} A^{0+} &=& -T' e^{i\gamma} -C' e^{i\gamma}
+P'_{tc} -~P'_{uc} e^{i\gamma} -~\pewp -\frac23 \pewcp ~,
\nn\\
A^{-+} &=& -T' e^{i\gamma} + P'_{tc} -P'_{uc}
e^{i\gamma} -\frac23 \pewcp ~, \nn\\
\sqrt{2} A^{00} &=& -C' e^{i\gamma} - P'_{tc} +P'_{uc}
e^{i\gamma} - \pewp -\frac13 \pewcp ~.
\eea
We have explicitly written the weak-phase dependence (including the
minus sign from $V_{tb}^* V_{ts}$ [in $P'_{tc}$]), so that the
diagrams contain both strong phases and the magnitudes of the
Cabibbo-Kobayashi-Maskawa (CKM) matrix elements. The amplitudes for
the CP-conjugate processes can be obtained from the above by changing
the sign of the weak phase $\gamma$.

$P'_{tc}$ and $P'_{uc}$ are defined as follows. There are three
gluonic penguin contributions $P_i$, where $i = u,c,t$ indicates the
identity of the quark in the loop. The full penguin amplitude is then
\bea
P &=& V_{ub}^* V_{us} P'_u + V_{cb}^* V_{cs} P'_c + V_{tb}^* V_{ts} P'_t \nn\\
&=& V_{ub}^* V_{us} (P'_u - P'_c) + V_{tb}^* V_{ts} (P'_t - P'_c) ~.
\label{penguin}
\eea
The second line arises due to the unitarity of the CKM matrix
($V_{ub}^* V_{us} + V_{cb}^* V_{cs} + V_{tb}^* V_{ts} = 0$), and we
define $P'_{tc} \equiv |V_{tb}^* V_{ts}| (P'_t - P'_c)$ and $P'_{uc}
\equiv |V_{ub}^* V_{us} | (P'_u - P'_c)$. Since $|V_{ub}^* V_{us} | =
O(\lambda^4)$ and $|V_{tb}^* V_{ts} | = O(\lambda^2)$, where $\lambda
= 0.22$ is the sine of the Cabibbo angle, $|P'_{uc}| \ll |P'_{tc}|$.

It has been shown \cite{EWPtree1, EWPtree2, EWPtree3} that, to a good
approximation, the diagrams $\pewp$ and $\pewcp$ can be related to
$T'$ and $C'$ within the SM using flavor SU(3) symmetry:
\bea
\label{EWPrels}
\pewp & \!\!=\!\! & {3\over 4} {c_9 + c_{10} \over c_1 + c_2} R (T' +
C') \!+\!  {3\over 4} {c_9 - c_{10} \over c_1 - c_2} R (T' - C')
~, \nn\\
\pewcp & \!\!=\!\! & {3\over 4} {c_9 + c_{10} \over c_1 + c_2} R (T' +
C') \!-\!  {3\over 4} {c_9 - c_{10} \over c_1 - c_2} R (T' - C')
~.
\eea
We refer to these as EWP-tree relations. Here, the $c_i$ are Wilson
coefficients \cite{BuraseffH} and $R \equiv \left\vert (V_{tb}^*
V_{ts})/(V_{ub}^* V_{us}) \right\vert = 49.1 \pm 1.0$
\cite{CKMfitter}. Now, $c_{10}/c_9 = c_2/c_1$ holds to a few
percent. In this limit, the above EWP-tree relations become
\beq
\label{approxEWPrels}
\pewp = {3\over 2} {c_9 \over c_1} R \, T' ~~,~~~~
\pewcp = {3\over 2} {c_9 \over c_1} R \, C' ~.
\eeq
Thus, $\pewp$ and $T'$ are roughly the same size, as are $\pewcp$ and
$C'$.

Taking the above information into account, the relative sizes of all
the $\btopik$ diagrams can be roughly estimated as\footnote{These
  estimates were first given in Ref.~\cite{GHLR2}, which predates the
  derivation of the EWP-tree relations.}
\beq
1 : |P'_{tc}| ~~,~~~~ {\cal O}({\bar\lambda}) : |T'|,~|\pewp|
~~,~~~~ {\cal O}({\bar\lambda}^2) : |C'|,~|P'_{uc}|,~|\pewcp|
~,
\label{ampsizes}
\eeq
where ${\bar\lambda} \sim 0.2$. 

\subsection{\bf \boldmath Naive $\btopik$ Puzzle}

Neglecting the diagrams of ${\cal O}({\bar\lambda}^2)$ in
Eq.~(\ref{ampsizes}), the $\btopik$ amplitudes become
\bea
\label{reducedamps}
A^{+0} &=& -P'_{tc} ~, \nn\\
\sqrt{2} A^{0+} &=& -T' e^{i\gamma} + P'_{tc} - \pewp ~, \nn\\
A^{-+} &=& -T' e^{i\gamma} + P'_{tc} ~, \nn\\
\sqrt{2} A^{00} &=& - P'_{tc} - \pewp ~.
\eea
With these amplitudes, consider the direct CP asymmetries of $B^+ \to
\pi^0 K^+$ and $\bd \to \pi^- K^+$. Such CP asymmetries are generated
by the interference of two amplitudes with nonzero relative weak and
strong phases. In both $A^{0+}$ and $A^{-+}$, $T'$-$P'_{tc}$
interference leads to a direct CP asymmetry. On the other hand, in
$A^{0+}$, $\pewp$ and $T'$ have the same strong phase ($\pewp \propto
T'$ [Eq.~(\ref{approxEWPrels})]), while $\pewp$ and $P'_{tc}$ have the
same weak phase ($=0$), so that $\pewp$ does not contribute to the
direct CP asymmetry. This means that we expect $A_{CP}(B^+ \to \pi^0
K^+) = A_{CP}(\bd \to \pi^- K^+)$.

The latest $\btopik$ measurements are shown in Table \ref{tab:data}.
Not only are $A_{CP}(B^+ \to \pi^0 K^+)$ and $A_{CP}(\bd \to \pi^-
K^+)$ not equal, they are of opposite sign!  Experimentally, we have
$(\Delta A_{CP})_{\rm exp} = (12.2 \pm 2.2) \%$. This differs from 0
by $5.5\sigma$.  This is the naive $\btopik$ puzzle.

\begin{table}[tbh]
\center
\begin{tabular}{|c|c|c|c|}
\hline
Mode & $BR[10^{-6}]$ & $A_{CP}$ & $S_{CP}$ \\ \hline
$B^+ \to \pi^+ K^0$ & $23.79 \pm 0.75$ & $-0.017 \pm 0.016$ & \\
\hline
$B^+ \to \pi^0 K^+$ & $12.94 \pm 0.52$ & $0.040 \pm 0.021$ & \\
\hline
$\bd \to \pi^- K^+$ & $19.57 \pm 0.53$ & $-0.082 \pm 0.006$ & \\
\hline
$\bd \to \pi^0 K^0$ & $9.93 \pm 0.49$ & $-0.01 \pm 0.10$ & $0.57 \pm 0.17$ \\
\hline
\end{tabular}
\caption{Branching ratios, direct CP asymmetries $A_{CP}$, and
  mixing-induced CP asymmetry $S_{CP}$ (if applicable) for the four
  $\btopik$ decay modes. The data are taken from Ref.~\cite{HFAG}.}
\label{tab:data}
\end{table}

\subsection{\bf Statistics}

As noted in the Introduction, a more accurate measure of the
(dis)agreement with the SM can be obtained by performing a fit to the
data. From here on, in looking for SM or NP explanations of the
$\btopik$ puzzle, we will use only fits. In such fits, the $\btopik$
amplitudes are expressed in terms of a certain number of unknown
theoretical parameters. In order to perform a fit, there must be fewer
theoretical unknowns than observables. We define $\chi^2$ as
\beq
\chi^2 = \sum_i \frac{ ({\cal O}_i^{th} - {\cal O}_i^{exp})^2 }{ (\Delta {\cal O}_i)^2 } ~,
\eeq
where ${\cal O}_i$ are the various observables used as constraints.
${\cal O}_i^{exp}$ and $\Delta {\cal O}_i$ are, respectively, the
experimentally-measured central values and errors.  ${\cal O}_i^{th}$
are the theoretical predictions for the observables, and are functions
of the unknown theoretical parameters.  We use the program {\tt
  MINUIT} \cite{James:1975dr,James:2004xla,James:1994vla} to find the
values of the unknowns that minimize the $\chi^2$.

At this point, it is useful to review some basic properties of
$\chi^2$ distributions in order to establish what constitutes a good
fit. The $\chi^2$ probability distribution depends on a single
parameter, $n$, which is the number of degrees of freedom (d.o.f.).
It is given by
\beq
P(\chi^2) = \frac{1}{2^{n/2} \, \Gamma(n/2)} \, (\chi^2)^{(n/2-1)} e^{-\chi^2/2} ~.
\eeq
For $n$ large, this becomes a normal distribution with central value
$n$ and standard deviation $\sqrt{2 n}$. That is, in this limit the
preferred value of $\chi^2/{\rm d.o.f.}$ is 1, with an error
$\sqrt{2/n}$. This says that, even if we have the correct underlying
theory, we still expect $\chi^2/{\rm d.o.f.} \simeq 1$, just due to
statistical fluctuations. For this reason, it is common to say that,
if we find $\chi^2_{\rm min}/{\rm d.o.f.} \simeq 1$ in a fit, it is
acceptable.

We stress that this only holds for $n$ large -- it is not justified to
apply the same criterion for small values of $n$.  One way to see this
is to compute the p-value. For $n$ large, the p-value corresponding to
$\chi^2/{\rm d.o.f.} = 1$ is 50\%. That is, a p-value of $\simeq 0.5$
constitutes an acceptable fit. However, here are the p-values
corresponding to $\chi^2/{\rm d.o.f.} = 1$ for smaller values of $n$:
\bea
n = 1 &:& {\hbox{p-value}} = 0.32 ~, \nn\\
n = 2 &:& {\hbox{p-value}} = 0.37 ~, \nn\\
n = 3 &:& {\hbox{p-value}} = 0.39 ~, \nn\\
n = 5 &:& {\hbox{p-value}} = 0.42 ~, \nn\\
n = 10 &:& {\hbox{p-value}} = 0.44 ~.
\eea
These reflect the fact that, for $n$ small, the central value of the
distribution is not at $\chi^2/{\rm d.o.f.} = 1$ -- it is at smaller
values. This shows that, if $n=1$, though $\chi^2/{\rm d.o.f.} = 1$ is
not a bad fit, it is still somewhat less than acceptable (which
corresponds to a p-value of $\simeq 0.5$). It also suggests that
p-values are easier than $\chi^2/{\rm d.o.f.}$ for judging the
goodness-of-fit when $n$ is small.

\subsection{\bf \boldmath True $\btopik$ Puzzle}

Taking into account the first EWP-tree relation of
Eq.~(\ref{approxEWPrels}), the amplitudes of Eq.~(\ref{reducedamps})
depend on four unknown parameters: the magnitudes $|T'|$ and
$|P'_{tc}|$, one relative strong phase, and the weak phase
$\gamma$. In addition, the indirect CP asymmetry in $\bd\to \pi^0K^0$,
$S_{CP}$, depends on the weak phase $\beta$. These parameters are
constrained by the $\btopik$ data of Table \ref{tab:data}, as well as
by the independent measurements of the weak phases \cite{CKMfitter}:
\beq
\beta = (21.85 \pm 0.68)^\circ ~~,~~~~ \gamma = (72.1 \pm 5.8)^\circ ~. 
\label{phasemeas}
\eeq
With more observables (11) than theoretical unknowns (5), a fit can be
performed. The results are shown in Table \ref{tab:smallfit}.
Unsurprisingly, we find a terrible fit: $\chi^2_{\rm min}/{\rm d.o.f.}
= 30.9/6$, corresponding to a p-value of $3.0 \times 10^{-5}$. This
can be considered the true $\btopik$ puzzle.

\begin{table}[tbh]
\center
\begin{tabular}{|c|c|}
\hline
\multicolumn{2}{|c|}{$\chi^2_{\rm min}/{\rm d.o.f.} = 30.9/6$,}\\
\multicolumn{2}{|c|}{p-value $=3.0 \times 10^{-5}$}\\
 \hline
Parameter & Best-fit value \\ 
\hline
$\gamma$ & $(67.2 \pm 4.7)^\circ$ \\
\hline
$\beta$ & $(21.80 \pm 0.68)^\circ$ \\
\hline
$|T'|$ & $7.0 \pm 1.4$ \\
\hline
$|P'_{tc}|$ & $50.5 \pm 0.6$ \\
\hline
$\delta_{P'_{tc}} - \delta_{T'}$ & $(-15.6 \pm 3.4)^\circ$ \\
\hline
\end{tabular}
\caption{$\chi^2_{\rm min}/{\rm d.o.f.}$ and best-fit values of unknown
  parameters in amplitudes of Eq.~(\ref{reducedamps}). Constraints:
  $\btopik$ data, measurements of $\beta$ and $\gamma$.}
\label{tab:smallfit}
\end{table}

\section{SM Fits}

In the previous section, we saw that the SM cannot account for the
$\btopik$ data if the small diagrams $C'$, $P'_{uc}$ and $\pewcp$ are
neglected. Thus, in order to test whether the data can be explained by
the SM, the small diagrams must be included in the fit. With the
EWP-tree relations of Eq.~(\ref{EWPrels}), there are only four
independent diagrams in the $\btopik$ amplitudes\footnote{The EWP-tree
  relations were first applied to $\btopik$ decays in
  Ref.~\cite{Imbeault:2003it}, which predates discussions of the
  $\btopik$ puzzle.}: $T'$, $C'$, $P'_{tc}$ and $P'_{uc}$. This
corresponds to 8 unknown parameters: four magnitudes of diagrams,
three relative strong phases, and the weak phase $\gamma$. (As before,
the value of $\beta$, which is required for $S_{CP}$, can be
constrained from independent measurements. This is done in all fits.)
We therefore have more observables (9) than theoretical unknowns (8),
so that a fit can be done. Additional constraints can come from the
independent measurement of $\gamma$ and/or theoretical input. In this
section we perform various SM fits in order to determine under what
circumstances the SM can explain the $\btopik$ puzzle.

Regarding the strong phases, these are mainly generated by QCD
rescattering from another diagram with the same CKM matrix elements.
As an example, take $P'_c$ in Eq.~(\ref{penguin}).  It can arise
directly via a gluonic penguin amplitude with a $c$ quark in the loop,
or it can be generated by rescattering from the tree operator ${\bar
  b} \to {\bar s} c {\bar c}$:
\beq
(P'_c)_{tot} = (P'_c)_{dir} + (P'_c)_{rescatt} ~.
\eeq
Only the rescattered contribution has a substantial strong phase. Now,
the tree diagram is much larger than the penguin diagram. But
rescattering comes with a cost: the rescattered penguin is only about
5-10\% as large as the tree. Indeed it is of the same order as
$(P'_c)_{dir}$. The net effect is that $(P'_c)_{tot}$ can have a
sizeable strong phase. On the other hand, the strong phase of $T'$ can
only arise due to self-rescattering. Since this self-rescattered
amplitude is only about 5-10\% as large as the original amplitude, the
strong phase of $T'$ is expected to be small. The bottom line is that
$\delta_{T'}$ should be small, while the other strong phases can be
large. In the fits we adopt the convention that $\delta_{T'} = 0$. (In
any case, only relative strong phases are measurable.)

\subsection{\bf \boldmath All diagrams free, constraint on $\gamma$ added}

In this fit, we keep all diagrams and allow their values to vary, but
constrain $\gamma$ by including the independent measurement of
Eq.~(\ref{phasemeas}) ($\beta$ is always constrained in this way.) The
results of the fit are given in Table \ref{tab:SMfitCTfree}. We see
that the p-value is only 17\%, which is below the 50\% required for an
acceptable fit.  More importantly, the best fit has $|C'/T'| = 0.75
\pm 0.32$, which is considerably larger than the estimate in
Eq.~(\ref{ampsizes}).

\begin{table}[tbh]
\center
\begin{tabular}{|c|c|}
\hline
\multicolumn{2}{|c|}{SM fit (2): $\chi^2/{\rm d.o.f.} = 3.5/2$,}\\
\multicolumn{2}{|c|}{~~~~~~~~~~~~~~~~~~ p-value $=0.17$}\\
 \hline
Parameter & Best-fit value \\ 
\hline
$\gamma$ & $(72.0 \pm 5.8)^\circ$ \\
\hline
$\beta$ & $(21.85 \pm 0.68)^\circ$ \\
\hline
$|T'|$ & $5.2 \pm 1.5$ \\
\hline
$|C'|$ & $3.9 \pm 1.2$ \\
\hline
$|P'_{tc}|$ & $50.7 \pm 0.9$ \\
\hline
$|P'_{uc}|$ & $1.1 \pm 2.4$ \\
\hline
$\delta_{C'}$ & $(209.8 \pm 21.3)^\circ$ \\
\hline
$\delta_{P'_{tc}}$ & $(-16.2 \pm 7.3)^\circ$ \\
\hline
$\delta_{P'_{uc}}$ & $(4.9 \pm 51.3)^\circ$ \\
\hline
\end{tabular}
\caption{$\chi^2_{\rm min}/{\rm d.o.f.}$ and best-fit values of unknown
  parameters in amplitudes of Eq.~(\ref{fulldiagrams}). Constraints:
  $\btopik$ data, measurements of $\beta$ and $\gamma$.}
\label{tab:SMfitCTfree}
\end{table}

Even though the fit is only fair, it is clear that the SM fit prefers
a large value of $|C'/T'|$. But this raises the question: what does
theory predict for $|C'/T'|$? 
\begin{enumerate}

\item In Ref.~\cite{QCDfinput}, $\btopik$ decays were analyzed in the
  context of QCD factorization (QCDf). The various NLO contributions
  were computed for three different values of the renormalization
  scale, $\mu = m_b/2$, $m_b$ and $2 m_b$. In all three cases it was
  found that $|C'/T'| \simeq 0.2$.

\item The NNLO corrections within QCDf have been considered in
  Refs.~\cite{QCDFNNLO1,QCDFNNLO2,QCDFNNLO3,QCDFNNLO4}.  Including
  these corrections, it is found \cite{privcomm} that $0.13 \le
  |C'/T'| \le 0.43$, with a central value of $|C'/T'| = 0.23$, very
  near its NLO value.

\item Ref.~\cite{Li:2009wba} does NNLO calculations within
  perturbative QCD (pQCD), and finds $|C'/T'| = 0.53$. We note that a
  range of values is not given, which suggests that this result is for
  a specific choice of the theoretical parameters. It is not clear
  what the smallest allowed value of $|C'/T'|$ is within pQCD.

\end{enumerate}
It therefore appears that $|C'/T'|$ can be as large as $\sim 0.5$. But
it may also be the case that $|C'/T'|$ is quite a bit smaller. In
light of this, below we repeat the SM fit, taking as theoretical input
$|C'/T'| = 0.2$ or 0.5.

There is one more thing. In Eq.~(\ref{ampsizes}), it is estimated that
$|P'_{uc}/P'_{tc}| = {\cal O}({\bar\lambda}^2)$. However, it has been
argued that $|P'_{uc}|$ is actually even smaller: $|P'_{uc}/P'_{tc}| =
{\cal O}({\bar\lambda}^3)$ \cite{Kim}, so that it can be neglected, to
a good approximation. Indeed, in Table \ref{tab:SMfitCTfree} $P'_{uc}$
is the smallest diagram. Therefore, from now on we will also add the
theoretical input $P'_{uc} = 0$. Note that this is the most favorable
assumption for the SM: it increases the d.o.f., and hence the p-value,
of a fit. If we find that a particular SM fit is poor, it would be
even worse had we allowed $P'_{uc}$ to vary.

\subsection{\bf \boldmath $|C'/T'| = 0.2$, $P'_{uc} = 0$, constraint on $\gamma$ added}

We now perform the same fit as above, but add the theoretical
constraints $|C'/T'| = 0.2$ and $P'_{uc} = 0$.  The results of the fit
are given in Table \ref{tab:SMfitCT0.2}. The situation is better than
in Table \ref{tab:smallfit}, but we still have a poor fit:
$\chi^2_{\rm min}/{\rm d.o.f.} = 12.1/5$, corresponding to a p-value
of 3\%. This demonstrates conclusively that, if $|C'/T'| = 0.2$, the
$\btopik$ puzzle cannot be explained by the SM.

\begin{table}[tbh]
\center
\begin{tabular}{|c|c|}
\hline
\multicolumn{2}{|c|}{$\chi^2_{\rm min}/{\rm d.o.f.} = 12.1/5$,}\\
\multicolumn{2}{|c|}{p-value $= 0.03$}\\
 \hline
Parameter & Best-fit value \\ 
\hline
$\gamma$ & $(67.2 \pm 4.6)^\circ$ \\
\hline
$\beta$ & $(21.80 \pm 0.68)^\circ$ \\
\hline
$|T'|$ & $7.9 \pm 1.2$ \\
\hline
$|P'_{tc}|$ & $50.7 \pm 0.6$ \\
\hline
$\delta_{P'_{tc}}$ & $(346.5 \pm 2.6)^\circ$ \\
\hline
$\delta_{C'}$ & $(253.1 \pm 23.5)^\circ$ \\
\hline
\end{tabular}
\caption{$\chi^2_{\rm min}/{\rm d.o.f.}$ and best-fit values of unknown
  parameters in amplitudes of Eq.~(\ref{fulldiagrams}). Constraints:
  $\btopik$ data, measurements of $\beta$ and $\gamma$, theoretical
  inputs $|C'/T'| = 0.2$, $P'_{uc} = 0$.}
\label{tab:SMfitCT0.2}
\end{table}

\subsection{\bf \boldmath $|C'/T'| = 0.5$, $P'_{uc} = 0$, constraints on $\gamma$ added}

We now perform the SM fit, but with the theoretical constraints
$|C'/T'| = 0.5$ and $P'_{uc} = 0$. The results of the fit are given in
Table \ref{tab:SMfitCT0.5}. We find $\chi^2_{\rm min}/{\rm d.o.f.} =
4.9/5$, for a p-value of 43\%, which is an acceptable fit.  We
therefore conclude that, if $|C'/T'| = 0.5$, there is no $\btopik$
puzzle -- the data can be explained by the SM.

\begin{table}[tbh]
\center
\begin{tabular}{|c|c|}
\hline
\multicolumn{2}{|c|}{$\chi^2_{\rm min}/{\rm d.o.f.} = 4.9/5$,}\\
\multicolumn{2}{|c|}{p-value $= 0.43$}\\
 \hline
Parameter & Best-fit value \\ 
\hline
$\gamma$ & $(70.6 \pm 5.3)^\circ$ \\
\hline
$\beta$ & $(21.82 \pm 0.68)^\circ$ \\
\hline
$|T'|$ & $6.2 \pm 0.9$ \\
\hline
$|P'_{tc}|$ & $50.5 \pm 0.5$ \\
\hline
$\delta_{P'_{tc}}$ & $(162.4 \pm 3.5)^\circ$ \\
\hline
$\delta_{C'}$ & $(42.8 \pm 18.1)^\circ$ \\
\hline
\end{tabular}
\caption{$\chi^2_{\rm min}/{\rm d.o.f.}$ and best-fit values of unknown
  parameters in amplitudes of Eq.~(\ref{fulldiagrams}). Constraints:
  $\btopik$ data, measurements of $\beta$ and $\gamma$, theoretical
  inputs $|C'/T'| = 0.5$, $P'_{uc} = 0$.}
\label{tab:SMfitCT0.5}
\end{table}

It is interesting to compare the results of this fit with those in
Table \ref{tab:SMfitCTfree}. By $\chi^2_{\rm min}$, this fit is worse,
since $(\chi^2_{\rm min})_{\rm Table~\ref{tab:SMfitCT0.5}} >
(\chi^2_{\rm min})_{\rm Table~\ref{tab:SMfitCTfree}}$. On the other
hand, by p-value it is better. The reason is that the number of
d.o.f.\ has changed, so that $(\chi^2_{\rm min}/{\rm d.o.f.})_{\rm
  Table~\ref{tab:SMfitCT0.5}} < (\chi^2_{\rm min}/{\rm d.o.f.})_{\rm
  Table~\ref{tab:SMfitCTfree}}$. This type of behavior often occurs
when the number of d.o.f.\ is small, making it difficult to judge
which of two fits is truly better.

Note that the strong phase $\delta_{C'}$ is produced mainly by
rescattering from $T'$.  Thus, for a large value of $|C'/T'|$,
$\delta_{C'}$ should be on the small side. And indeed, from Table
\ref{tab:SMfitCT0.5}, we see that $\delta_{C'}$ is not that large.

Now, in this fit $\gamma$ is constrained by its independently-measured
value.  However, this is not necessary -- $\gamma$ can be treated as
an unknown parameter and extracted from the measurements of $\btopik$
decays. Indeed, if the SM explains the $\btopik$ data, we would expect
the extracted value of $\gamma$ to be the same as that measured in
tree-level decays. We investigate this in the following subsection.

\subsection{\bf \boldmath $|C'/T'| = 0.5$, $P'_{uc} = 0$, $\gamma$ free}

Here we repeat the fit of the previous subsection, but remove the
constraint from the independent measurement of $\gamma$. The results
are given in Table \ref{tab:SMfitgammafree}. By $\chi^2_{\rm min}$,
this fit is better than that of Table \ref{tab:SMfitCT0.5}. However,
by p-value it is worse. This is a result of the different d.o.f.\ in
the two fits. More importantly, it is found that $\gamma = (51.2 \pm
5.1)^\circ$, which deviates from its measured value of $(72.1 \pm
5.8)^\circ$ by $2.7\sigma$. So this is a reason not to be entirely
satisfied that the SM explains the $\btopik$ puzzle, even if $|C'/T'|
= 0.5$.

\begin{table}[tbh]
\center
\begin{tabular}{|c|c|}
\hline
\multicolumn{2}{|c|}{$\chi^2_{\rm min}/{\rm d.o.f.} = 4.3/4$,}\\
\multicolumn{2}{|c|}{p-value $= 0.36$}\\
 \hline
Parameter & Best-fit value \\ 
\hline
$\gamma$ & $(51.2 \pm 5.1)^\circ$ \\
\hline
$\beta$ & $(21.78 \pm 0.68)^\circ$ \\
\hline
$|T'|$ & $10.1 \pm 3.4$ \\
\hline
$|P'_{tc}|$ & $51.8 \pm 1.0$ \\
\hline
$\delta_{P'_{tc}}$ & $(168.6 \pm 4.6)^\circ$ \\
\hline
$\delta_{C'}$ & $(131.2 \pm 24.7)^\circ$ \\
\hline
\end{tabular}
\caption{$\chi^2_{\rm min}/{\rm d.o.f.}$ and best-fit values of
  unknown parameters in amplitudes of Eq.~(\ref{fulldiagrams}).
  Constraints: $\btopik$ data, measurement of $\beta$, theoretical
  inputs $|C'/T'| = 0.5$, $P'_{uc} = 0$.}
\label{tab:SMfitgammafree}
\end{table}

\subsection{\bf Summary}

If $|C'/T'| = 0.2$, we have found that the SM cannot explain the
$\btopik$ puzzle. Even in the most optimistic scenario, where
$P'_{uc}$ is neglected, the p-value is only 3\%, constituting a poor
fit. On the other hand, if $|C'/T'| = 0.5$ and $P'_{uc} = 0$, the
p-value is 43\%, which is acceptable.  There is still an aspect of
this fit that is not entirely satisfactory, but on the whole we find
that, in this case, the SM can explain the $\btopik$ data.

We therefore conclude that it is the size of $|C'/T'|$ that determines
whether or not there is truly a $\btopik$ puzzle. If $|C'/T'| = 0.5$,
which is theoretically its maximally-allowed value, the SM can explain
the $\btopik$ data. However, if $|C'/T'| = 0.2$, which is towards the
lower end of its theoretically-allowed range, then the SM cannot
explain the data, and NP is required. In this case, it is natural to
investigate what type of NP is required. We do this in the next
section.

\section{NP Fits}

In this section we assume $|C'/T'| = 0.2$ and examine whether the
$\btopik$ puzzle can be explained with the addition of new physics.

\subsection{\bf Model-independent formalism}
\label{ModelindepNP}

In the general approach of Ref.~\cite{DLNP}, the NP operators that
contribute to the $\btopik$ amplitudes take the form ${\cal
  O}_{NP}^{ij,q} \sim {\bar s} \Gamma_i b \, {\bar q} \Gamma_j q$ ($q
= u,d$), where $\Gamma_{i,j}$ represent Lorentz structures, and color
indices are suppressed. The NP contributions to $\btopik$ are encoded
in the matrix elements $\bra{\pi K} {\cal O}_{NP}^{ij,q} \ket{B}$. In
general, each matrix element has its own NP weak and strong phases.

Now, above we noted that strong phases are basically generated by QCD
rescattering from diagrams with the same CKM matrix elements. We then
argued that the strong phase of $T'$ is expected to be very small
since it is due to self-rescattering. For the same reason, all NP
strong phases are also small, and can be neglected. In this case, many
NP matrix elements can be combined into a single NP amplitude, with a
single weak phase:
\beq
\sum \bra{\pi K} {\cal O}_{NP}^{ij,q} \ket{B} = \ANPq
e^{i \Phi_q} ~.
\eeq
Here the strong phase is zero. There are two classes of such NP
amplitudes, differing only in their color structure: ${\bar s}_\alpha
\Gamma_i b_\alpha \, {\bar q}_\beta \Gamma_j q_\beta$ and ${\bar
  s}_\alpha \Gamma_i b_\beta \, {\bar q}_\beta \Gamma_j q_\alpha$ ($q
= u,d$). They are denoted $\ApNPqph$ and $\ApNPCqph$, respectively
\cite{BNPmethods}. Here, $\Phi'_q$ and $\Phi_q^{\prime {C}}$ are the
NP weak phases. In general, ${\cal A}^{\prime,q} \ne {\cal A}^{\prime
  {C}, q}$ and $\Phi'_q \ne \Phi_q^{\prime {C}}$. Note that, despite
the ``color-suppressed'' index $C$, the matrix elements $\ApNPCqph$
are not necessarily smaller than $\ApNPqph$.

There are therefore four NP matrix elements that contribute to
$\btopik$ decays. However, only three combinations appear in the
amplitudes: $\ApNPcomb \equiv - \ApNPuph + \ApNPdph$, $\ApNPCuph$, and
$\ApNPCdph$ \cite{BNPmethods}. The $\btopik$ amplitudes can now be
written in terms of the SM diagrams and these NP matrix elements. Here
we neglect the small SM diagram $P'_{uc}$:
\bea
\label{BpiKNPamps}
A^{+0} &=& -P'_{tc} -\frac13 \pewcp + \ApNPCdph ~, \nn\\
\sqrt{2} A^{0+} &=& P'_{tc} - T' \, e^{i\gamma} - \pewp -C' \, e^{i\gamma} -\frac23 \pewcp + \ApNPcomb - \ApNPCuph ~, \nn\\
A^{-+} &=& P'_{tc} - T' \, e^{i\gamma} -\frac23 \pewcp - \ApNPCuph ~, \nn\\
\sqrt{2} A^{00} &=& -P'_{tc} - \pewp -C' \, e^{i\gamma} -\frac13 \pewcp + \ApNPcomb + \ApNPCdph ~.
\eea

In Ref.~\cite{Baek:2009pa}, a different set of NP operators is defined:
\bea
\pewpnp \, e^{i \Phi'_{EW}} & \equiv & \ApNPuph - \ApNPdph ~, \nn\\
P'_{NP} \, e^{i \Phi'_{P}} & \equiv & \frac13 \ApNPCuph\ + \frac23 \ApNPCdph~, \nn\\
\pewcpnp \, e^{i \Phi^{\prime {C}}_{EW}} & \equiv & \ApNPCuph\ - \ApNPCdph ~.
\label{NPPoperators}
\eea
In order, these imply the inclusion of NP in the color-allowed
electroweak penguin, the gluonic penguin, and the color-suppressed
electroweak penguin amplitudes. Of course, the two sets of NP
operators, $\{ \ApNPcomb, \ApNPCuph, \ApNPCdph \}$ and $\{ \pewpnp,
P'_{NP}, \pewcpnp \}$, are equivalent. Depending on the situation, one
set or the other may be used.

In the most general case, there are three independent NP operators
that contribute to $\btopik$ decays. Although their strong phases are
negligible, they each have their own weak phase. There are therefore
12 parameters in the $\btopik$ amplitudes: 6 magnitudes of diagrams, 2
relative strong phases, 3 NP weak phases, and the CKM phase $\gamma$.
As before, the CKM phase $\beta$ also contributes to the indirect CP
asymmetry in $\bd\to \pi^0K^0$, so that there are a total of 13
unknown parameters. However, there are only 12 constraints: the 9
$\btopik$ observables, the independent measurements of $\beta$ and
$\gamma$, and the theoretical input $|C'/T'| = 0.2$. With more
unknowns than constraints, a fit cannot be done. However, this may be
improved in the context of a specific model. There may be fewer
unknown parameters, or they may not all be independent. We examine
this possibility in the following subsections.

\subsection{\bf \boldmath $Z'$ models}

For $\btopik$, the relevant decay is $\bsqq$ ($q=u,d$). This can occur
via the tree-level exchange of a $Z'$ that has a flavor-changing
coupling to ${\bar s}b$ and also couples to ${\bar q}q$. The exact
form of the ${\bar s}b Z'$ coupling is unimportant, and for the
light quarks, all four currents $g_{L(R)}^{qq} \, {\bar q}
\gamma^{\mu} P_{L(R)} q$ are possible. Assuming the weak and mass
eigenstates are the same (i.e., we neglect the off-diagonal terms in
the CKM matrix), the only thing that is certain is that $g_L^{dd} =
g_L^{uu}$ due to $SU(2)_L$ symmetry.

Still, how the $Z'$ couples to ${\bar d}d$ and ${\bar u}u$ has direct
consequences for the NP operators:
\begin{enumerate}

\item Suppose that the $Z'$ couples only to left-handed ${\bar d}d$
  and ${\bar u}u$, i.e., $g_R^{dd} = g_R^{uu} = 0$. Since $g_L^{dd} =
  g_L^{uu}$, the NP operators $\ApNPdph$ and $\ApNPuph$ are equal, as
  are $\ApNPCdph$ and $\ApNPCuph$. In this case, $\pewpnp = \pewcpnp =
  0$ in Eq.~(\ref{NPPoperators}); the only nonzero NP operator is
  $P'_{NP}$. This also holds if the $Z'$ couples vectorially to ${\bar
    d}d$ and ${\bar u}u$, in which case $g_L^{dd} = g_R^{dd} =
  g_L^{uu} = g_R^{uu}$.

\item On the other hand, if $g_R^{dd}$ and $g_R^{uu}$ are nonzero, but
  $g_R^{uu} = -2 g_R^{dd}$, then $P'_{NP} = 0$, but $\pewpnp$ and
  $\pewcpnp$ are nonzero.

\item Alternatively, if only $g_R^{dd}$ is nonzero, $\ApNPcomb$ and
  $\ApNPCdph$ are nonzero (equivalently, $\pewpnp$ is nonzero and
  $P'_{NP} \, e^{i \Phi'_{P}} = -(2/3) \pewcpnp \, e^{i \Phi^{\prime
    {C}}_{EW}}$).

\item Similarly, if only $g_R^{uu}$ is nonzero, $\ApNPcomb$ and
  $\ApNPCuph$ are nonzero (equivalently, $\pewpnp$ is nonzero and
  $P'_{NP} \, e^{i \Phi'_{P}} = (1/3) \pewcpnp \, e^{i \Phi^{\prime
    {C}}_{EW}}$).

\item For all other choices of couplings, all three NP operators are
  nonzero.

\end{enumerate}
The point is that, although we consider a specific NP model, it has
many variations, so that its study includes a number of different NP
scenarios. Below we consider each of the cases above (which are
identified by their number in the list).

Note also that, for this particular kind of NP, we do naively expect the
color-suppressed operators to be smaller than the color-allowed ones.

The four-fermion operator corresponding to $\bsqq$ is proportional to
$g_L^{bs} g_{L(R)}^{qq}$.  Note that $g_{L(R)}^{qq}$ must be real,
since the light-quark current ${\bar q} \gamma^{\mu} P_{L(R)} q$ is
self-conjugate. However, $g_L^{bs}$ can be complex, i.e., it can
contain a weak phase. This is the only source of CP violation in the
model and, as it appears in all NP operators, the weak phases of all
operators are equal, i.e., there is only a single NP weak phase. That
is, $\Phi'_d = \Phi'_u = \Phi_d^{\prime C} = \Phi_u^{\prime C} \equiv
\Phi'$ for $\{ \ApNPcomb, \ApNPCuph, \ApNPCdph \}$ and $\Phi'_{EW} =
\Phi'_{P} = \Phi^{\prime {C}}_{EW} \equiv \Phi'$ for $\{ \pewpnp,
P'_{NP}, \pewcpnp \}$.

Now, $Z'$ models with a flavor-changing coupling to ${\bar s} b$ also
contribute to $\bs$-$\bsbar$ mixing.  The larger $g_L^{bs}$ is, the
more $Z'$ models contribute to -- and receive constraints from -- this
mixing. In particular, the phase of $\bs$-$\bsbar$ mixing has been
measured to be quite small: $\varphi_s^{c{\bar c}s} = -0.030 \pm
0.033$ \cite{HFAG}. Thus, if $g_L^{bs}$ is big, its phase, which is
the NP weak phase $\Phi'$ in $\btopik$, must be small. $\Phi'$ can be
large only if $g_L^{bs}$ is small, which means $g_{L(R)}^{qq}$ is
big. (This type of argument was first made in
Ref.~\cite{Alok:2017jgr}.)

\subsubsection{\bf Case (5): all three NP operators nonzero}

We begin with the general case, in which all three NP operators are
nonzero. Since all three NP weak phases are equal in the $Z'$ model,
there are 11 unknown parameters. However, there are 12 constraints --
the 9 $\btopik$ observables, the independent measurements of $\beta$
and $\gamma$, and the theoretical input $|C'/T'| = 0.2$ -- so a fit
can be done.

The results are shown in Table \ref{tab:NPfit3NPops} (left-hand
table). With $\chi^2/{\rm d.o.f.} = 0.41/1$ and a p-value of 52\%,
this is an excellent fit. However, there is a serious problem: the
best fit has $\pewcpnp/\pewpnp = 16$, whereas, in the $Z'$ model, the
color-suppressed NP operators are expected to be smaller than the
color-allowed ones. We therefore conclude that this result cannot
arise within a $Z'$ NP model.

\begin{table}[tbh]
\centering
\begin{tabular}{|c|c|}
\hline
\multicolumn{2}{|c|}{NP fit (5): $\chi^2/{\rm d.o.f.} = 0.41/1$,}\\
\multicolumn{2}{|c|}{~~~~~~~~~~~~~~~~~~ p-value $=0.52$}\\
 \hline
Parameter & Best-fit value \\ 
\hline
$\gamma$ & $(72.1 \pm 5.8)^\circ$ \\
\hline
$\beta$ & $(21.85 \pm 0.68)^\circ$ \\
\hline
$\Phi'$ & $(77.5 \pm 20.1)^\circ$ \\
\hline
$|T'|$ & $19.8 \pm 4.3$ \\
\hline
$|P'_{tc}|$ & $49.7 \pm 0.9$ \\
\hline
$P'_{NP}$ & $7.4 \pm 1.4$ \\
\hline
$\pewpnp$ & $1.3 \pm 8.6$ \\
\hline
$\pewcpnp$ & $20.9 \pm 4.2$ \\
\hline
$\delta_{P'_{tc}}$ & $(257.4 \pm 7.1)^\circ$ \\
\hline
$\delta_{C'}$ & $(91.6 \pm 126.0)^\circ$ \\
\hline
\end{tabular}
~~~~~~~~~~
\begin{tabular}{|c|c|}
\hline
\multicolumn{2}{|c|}{NP fit (5): $\chi^2/{\rm d.o.f.} = 1.85/2$,}\\
\multicolumn{2}{|c|}{~~~~~~~~~~~~~~~~~~ p-value $=0.4$}\\
 \hline
Parameter & Best-fit value \\ 
\hline
$\gamma$ & $(70.8 \pm 5.3)^\circ$ \\
\hline
$\beta$ & $(21.80 \pm 0.68)^\circ$ \\
\hline
$\Phi'$ & $(34.7 \pm 12.8)^\circ$ \\
\hline
$|T'|$ & $17.0 \pm 8.6$ \\
\hline
$|P'_{tc}|$ & $60.8 \pm 10.5$ \\
\hline
$P'_{NP}$ & $14.9 \pm 13.7$ \\
\hline
$\pewpnp$ & $11.6 \pm 8.5$ \\
\hline
$\pewcpnp$ & $3.5 \pm 2.7$ \\
\hline
$\delta_{P'_{tc}}$ & $(177.9 \pm 2.1)^\circ$ \\
\hline
$\delta_{C'}$ & $(64.1 \pm 58.3)^\circ$ \\
\hline
\end{tabular}
\caption{$\chi^2_{\rm min}/{\rm d.o.f.}$ and best-fit values of
  unknown parameters for the $Z'$ model where all three NP operators
  are present in $\btopik$. Left-hand table has constraints: $\btopik$
  data, measurements of $\beta$ and $\gamma$, $|C'/T'| = 0.2$.
  Right-hand table has constraints: $\btopik$ data, measurements of
  $\beta$ and $\gamma$, $|C'/T'| = 0.2$, and $|\pewcpnp/\pewpnp| =
  0.3$.}
\label{tab:NPfit3NPops}
\end{table}

To focus on a possibility consistent with a $Z'$ NP model, we impose
an additional (theoretical) constraint: $|\pewcpnp/\pewpnp| = 0.3$.
The results of this fit are shown in Table \ref{tab:NPfit3NPops}
(right-hand table). Here $\chi^2/{\rm d.o.f.} = 1.85/2$, for a p-value
of 40\%, which is a good fit. Of course, this good p-value is a
consequence of the fact that, with the constraint, the d.o.f.\ has
increased from 1 to 2. If we consider instead d.o.f $=1$ (which
essentially corresponds to the region in the space of the fit of Table
\ref{tab:NPfit3NPops} (left-hand table) with $|\pewcpnp/\pewpnp| =
0.3$), the p-value is 17\%. Although far from an excellent fit, it is
still better than that of the SM.

In the above fits, all NP operators are allowed. In the following, we
examine whether a better fit can be found if the model contains only a
subset of the operators. Clearly the minimum $\chi^2$ will be larger
than that found above, but since the d.o.f.\ will also be larger, a
larger p-value may be found, indicative of a better fit.

\subsubsection{\bf \boldmath Case (1): only $P'_{NP}$ nonzero}
\label{PNPnonzero}

The case where only $P'_{NP}$ is nonzero arises when the $Z'$
couplings to right-handed $d$ and $u$ quarks obey $g_R^{dd} =
g_R^{uu}$ (these can both vanish or be nonzero). Since $g_L^{dd} =
g_L^{uu}$ by weak isospin invariance, $\ApNPdph = \ApNPuph$ and
$\ApNPCdph = \ApNPCuph$, so that $\pewpnp = \pewcpnp = 0$. In
model-building terms, this corresponds to the case where the $Z'$
couples only to left-handed ${\bar d}d$ and ${\bar u}u$, or where it
couples vectorially to these quark pairs.

Assuming that only $P'_{NP}$ is nonzero, we note that all four
$\btopik$ amplitudes of Eq.~(\ref{BpiKNPamps}) contain the following
combination: $P'_{tc} - P'_{NP} e^{i \Phi'_P}$. Writing $P'_{tc} =
|P'_{tc}| e^{i \delta_{P'_{tc}}}$, this contains the four quantities
$|P'_{tc}|$, $\delta_{P'_{tc}}$, $P'_{NP}$ and $\Phi'_P$.  However,
here these are not all independent. One can see this by noting that
the combinations that appear in $B$ and ${\bar B}$ decays are:
\bea
|P'_{tc}| e^{i \delta_{P'_{tc}}} - P'_{NP} e^{i \Phi'_P} & \equiv & z ~, \nn\\
|P'_{tc}| e^{i \delta_{P'_{tc}}} - P'_{NP} e^{-i \Phi'_P} & \equiv & z' ~.
\label{zz'defs}
\eea
$z$ and $z'$ are complex numbers; their four real and imaginary parts
can be written in terms of the four theoretical parameters.  However,
it is clear from the above expressions that ${\rm Re}(z) = {\rm
  Re}(z')$.

In order to take this into account, we use ${\rm Re}(z)$, ${\rm
  Im}(z)$ and ${\rm Im}(z')$ as unknown parameters in the fit. The
results are shown in Table \ref{tab:NPfitP'only}.

\begin{table}[tbh]
\center
\begin{tabular}{|c|c|}
\hline
\multicolumn{2}{|c|}{NP fit (1): $\chi^2/{\rm d.o.f.} = 3.5/1$,}\\
\multicolumn{2}{|c|}{~~~~~~~~~~~~~~~~~~ p-value $=0.06$}\\
 \hline
Parameter & Best-fit value \\ 
\hline
$\gamma$ & $(72.0 \pm 5.9)^\circ$ \\
\hline
$\beta$ & $(21.85 \pm 0.68)^\circ$ \\
\hline
$|T'|$ & $5.2 \pm 1.5$ \\
\hline
${\rm Re}(z)$ & $-48.3 \pm 2.0$ \\
\hline
${\rm Im}(z)$ & $15.4 \pm 6.9$ \\
\hline
${\rm Im}(z')$ & $13.0 \pm 9.0$ \\
\hline
$|C'|$ & $3.9 \pm 1.2$ \\
\hline
$|P'_{uc}|$ & $0.2 \pm 10.6$ \\
\hline
$\delta_{C'}$ & $(29.8 \pm 21.3)^\circ$ \\
\hline
$\delta_{P'_{uc}}$ & $(333 \pm 360)^\circ$ \\
\hline
\end{tabular}
\caption{$\chi^2_{\rm min}/{\rm d.o.f.}$ and best-fit values of
  unknown parameters [$z$ and $z'$ are defined in Eq.~(\ref{zz'defs})]
  for the $Z'$ model where NP is present in $\btopik$, but only
  $P'_{NP}$ is nonzero.  Constraints: $\btopik$ data, measurements of
  $\beta$ and $\gamma$.}
\label{tab:NPfitP'only}
\end{table}

With $\chi^2_{\rm min}/{\rm d.o.f.}  = 3.5/1$ and a p-value of 6\%,
this is a poor fit, despite the addition of NP. But we can understand
what's going on here. We have used ${\rm Re}(z)$, ${\rm Im}(z)$ and
${\rm Im}(z')$ as unknown parameters. Referring to
Eq.~(\ref{zz'defs}), we see that, if $P'_{NP} = 0$, ${\rm Im}(z) =
{\rm Im}(z')$. But this is essentially what is found in Table
\ref{tab:NPfitP'only}. That is, with this particular type of NP, we
cannot do better than the SM. (This was also the conclusion of
Ref.~\cite{Baek:2009pa}.) Indeed, comparing with Table
\ref{tab:SMfitCTfree}, we see that the results are very similar. In
particular, the $\chi^2_{\rm min}$ is identical (the p-values are
different due to the different d.o.f.).

We therefore see that the only way for this NP to improve on the SM is
if the $Z'$ couplings to right-handed $d$ and $u$ quarks obey
$g_R^{dd} \ne g_R^{uu}$.

\subsubsection{\bf \boldmath Cases (3), (2), (4)}
\label{Z'cases324}

We begin with case (3), where only $g_R^{dd}$ is nonzero. Now
$\pewpnp$ is nonzero and $P'_{NP} = -(2/3) \pewcpnp$). We also impose
the constraint $|\pewcpnp/\pewpnp| = 0.3$. The results of the fit for
this case are shown in Table \ref{tab:NPfitcase3}. Here the p-value is
30\%, which is not bad (and is far better than that of the SM).

\begin{table}[tbh]
\center
\begin{tabular}{|c|c|}
\hline
\multicolumn{2}{|c|}{NP fit (3): $\chi^2/{\rm d.o.f.} = 3.67/3$,}\\
\multicolumn{2}{|c|}{~~~~~~~~~~~~~~~~~~ p-value $=0.30$}\\
 \hline
Parameter & Best-fit value \\ 
\hline
$\gamma$ & $(68.1 \pm 3.7)^\circ$ \\
\hline
$\beta$ & $(21.80 \pm 0.68)^\circ$ \\
\hline
$\Phi'$ & $(29.0 \pm 12.4)^\circ$ \\
\hline
$|T'|$ & $22.1 \pm 10.7$ \\
\hline
$|P'_{tc}|$ & $53.3 \pm 2.0$ \\
\hline
$\pewpnp$ & $14.8 \pm 9.3$ \\
\hline
$\pewcpnp$ & $4.2 \pm 2.9$ \\
\hline
$\delta_{P'_{tc}}$ & $(176.5 \pm 2.4)^\circ$ \\
\hline
$\delta_{C'}$ & $(42.5 \pm 28.9)^\circ$ \\
\hline
\end{tabular}
\caption{$\chi^2_{\rm min}/{\rm d.o.f.}$ and best-fit values of
  unknown parameters for the $Z'$ model where only $g_R^{dd}$ is
  nonzero. Constraints: $\btopik$ data, measurements of $\beta$ and
  $\gamma$, $|C'/T'| = 0.2$, $|\pewcpnp/\pewpnp| = 0.3$.}
\label{tab:NPfitcase3}
\end{table}

One important question is: what values of the $Z'$ mass and couplings
are required to explain the $\btopik$ puzzle? This can be deduced from
Table \ref{tab:NPfitcase3}. The SM $T'$ diagram involves the
tree-level decay ${\bar b} \to {\bar u} W^{+*} (\to u {\bar s} =
K^+)$.  The NP $\pewpnp$ diagram looks very similar -- we have the
tree-level decay ${\bar b} \to {\bar s} Z^{\prime *} (\to d {\bar d} =
\pi^0/\sqrt{2})$. Within factorization, the SM and NP diagrams involve
$A_{\pi K} \equiv F_0^{B \to \pi}(0) f_K$ and $A_{K \pi} \equiv F_0^{B
  \to K}(0) f_\pi$, respectively, where $F_0^{B \to K,\pi}(0)$ are
form factors and $f_{\pi,K}$ are decay constants. The hadronic factors
are similar in size: $|A_{K \pi}/A_{\pi K}| = 0.9 \pm 0.1$
\cite{QCDfinput}.  Taking central values, we have
\bea
\label{constraint1}
&& \left\vert \frac{\pewpnp}{T'} \right\vert \simeq 
\frac{A_{K \pi} |g_L^{bs} g_R^{dd}|/M_{Z'}^2}{A_{\pi K} (G_F/\sqrt{2})|V_{ub}^* V_{us}|} = \frac{14.8}{22.1} \nn\\
&\Longrightarrow& \frac{|g_L^{bs} g_R^{dd}|}{M_{Z'}^2} = 5.6 \times 10^{-3} ~ {\rm TeV}^{-2} ~.
\eea
This particular $Z'$ has no couplings to leptons, i.e., it is
leptophobic. The experimental limits on such $Z'$ bosons are very
weak. However, suppose they were stronger: say $M_{Z'} > 1$ TeV is
required.  The above constraint can still be satisfied with such
values of $M_{Z'}$, while keeping perturbative couplings. 

The only additional constraint comes from $\bs$-$\bsbar$ mixing. The
formalism is described in Ref.~\cite{Alok:2017jgr}, to which we refer
the reader for details. Briefly, in the presence of SM and NP
contributions, $\bs$-$\bsbar$ mixing is due to the operator
\beq
N C_{VLL} \, ({\bar s}_L \gamma^\mu b_L)\,({\bar s}_L \gamma_\mu b_L) ~,
\eeq
where
\beq
\label{BsmixSMNP}
N C_{VLL} \equiv |N C_{VLL}^{\rm SM}| e^{-2 i \beta_s}  + \frac{(g_L^{bs})^2}{2 M_{Z'}^2} ~.
\eeq
(Recall that $g_L^{bs} = |g_L^{bs}| e^{i\Phi'}$.) The $\bs$-$\bsbar$
mixing parameters are given by
\bea
\Delta M_s  &=& \frac{2}{3} m_{B_s} f_{B_s}^2 \hat B_{B_s}  \left | N C_{VLL} \right| ~, \nn\\
\varphi_s &=& {\rm arg}(N C_{VLL}) ~.
\eea
Here, $f_{B_s} \sqrt{\hat B_{B_s}}=270\pm 16$ MeV. The experimental
measurements of the mixing parameters yield \cite{HFAG}
\bea
\Delta M_s^{\rm exp} & = & 17.757 \pm 0.021 ~{\rm ps}^{-1} ~, \nn\\
\varphi_s^{c{\bar c}s} &=& -0.030 \pm 0.033 ~,
\eea
while the SM predictions are
\bea
\Delta M_s^{\rm SM} &=& \frac{2}{3} m_{B_s} f_{B_s}^2 \hat B_{B_s}
|N C_{VLL}^{\rm SM} | = (17.9 \pm 2.4)~{\rm ps}^{-1} ~, \nn\\
\varphi_s^{c{\bar cs,{\rm SM}}} &=& -2 \beta_s = -0.03704 \pm 0.00064 ~.
\eea

With all of this, we can obtain an upper bound on the NP contribution.
From Eq.~(\ref{BsmixSMNP}), we have
\beq
\frac{|g_L^{bs}|^2}{2 M_{Z'}^2} \le \sqrt{ |N C_{VLL}|^2 + |N C_{VLL}^{\rm SM}|^2 
- 2 |N C_{VLL}| |N C_{VLL}^{\rm SM}| \cos( \varphi_s^{c{\bar c}s} - \varphi_s^{c{\bar cs,{\rm SM}}} )} ~.
\eeq
We allow the experimental quantities to vary by $\pm 2\sigma$, while
$f_{B_s} \sqrt{\hat B_{B_s}}$ varies within its theoretical
range. This leads to
\beq
\label{constraint2}
\frac{|g_L^{bs}|^2}{M_{Z'}^2} \le 3.56 \times 10^{-5} ~{\rm TeV}^{-2} 
~~~~ \Longrightarrow ~~~~ M_{Z'} \ge |g_L^{bs}| \times 168 ~{\rm TeV} ~.
\eeq
(A similar result can be found in Ref.~\cite{ASBsmix}.) Combining this
with Eq.~(\ref{constraint1}), we have
\beq
\label{NPcoupratio}
\left\vert \frac{g_L^{bs}}{g_R^{dd}} \right\vert \le 6.4 \times 10^{-3} ~.
\eeq
With this condition, all the constraints can be satisfied. Therefore
this $Z'$ can indeed explain the $\btopik$ puzzle.

The results for cases (2) ($g_R^{dd}$ and $g_R^{uu}$ are nonzero, but
$g_R^{uu} = -2 g_R^{dd}$) and (4) (only $g_R^{uu}$ is nonzero) are
similar. The fit for case (2) has a p-value of 28\%, while that for
case (4) has p-value $=26\%$.

We therefore conclude that a $Z'$ model can explain the $\btopik$
puzzle, but it is necessary that the $Z'$ couple to right-handed $d$
and/or $u$ quarks, with $g_R^{dd} \ne g_R^{uu}$.

\subsubsection{\bf\boldmath $Z'$ $\bsmumu$ models and $\bsqq$}

As noted in the introduction, there are currently several measurements
that disagree with the predictions of the SM. These include (i) $R_K$
(LHCb \cite{RKexpt}) and (ii) $R_{K^*}$ (LHCb \cite{RK*expt}), (iii)
the angular distribution of $B \to K^* \mu^+\mu^-$ (LHCb
\cite{BK*mumuLHCb1, BK*mumuLHCb2}, Belle \cite{BK*mumuBelle}, ATLAS
\cite{BK*mumuATLAS} and CMS \cite{BK*mumuCMS}), especially the
observable $P'_5$ \cite{P'5}, and (iv) the branching fraction and
angular distribution of $\bs \to \phi \mu^+ \mu^-$ (LHCb
\cite{BsphimumuLHCb1, BsphimumuLHCb2}). The discrepancies in $R_K$ and
$R_{K^*}$ are quite clean and are at the level of 2.2-2.6$\sigma$. On
the other hand, the discrepancies in $B \to K^* \mu^+\mu^-$ and $\bs
\to \phi \mu^+ \mu^-$ have some amount of theoretical input. Depending
on how one treats the hadronic uncertainties, the disagreement with
the SM is in the 2.5-4$\sigma$ range.

Of these measurements, the most recent is that of $R_{K^*}$.
Following its announcement, a number of papers appeared
\cite{Capdevila:2017bsm, Altmannshofer:2017yso, DAmico:2017mtc,
  Hiller:2017bzc, Geng:2017svp, Ciuchini:2017mik, Celis:2017doq,
  Ghosh:2017ber, Alok:2017sui, Wang:2017mrd, Datta:2017ezo} computing
the size of the discrepancy with the SM, and determining the general
properties of the NP required to explain the results. Combining
constraints from all measurements, the general consensus is that there
is indeed a significant disagreement with the SM, somewhere in the
range of 4-6$\sigma$ (this large range is due to the fact that
different groups deal with the theoretical uncertainties in different
ways). In order to account for all measurements, the most probable
explanation is that the NP primarily affects $\bsmumu$ transitions.

Arguably the simplest NP explanation is that the contribution to
$\bsmumu$ arises due to the tree-level exchange of a $Z'$ boson. Here
the $Z'$ has a flavor-changing coupling to ${\bar s} b$, and it also
couples to ${\bar\mu} \mu$. Many models of this type have been
proposed to explain the data. In some of these models, the $Z'$ also
has couplings to ${\bar u} u$ and/or ${\bar d} d$, so that it could
potentially furnish an explanation of the $\btopik$ puzzle. If so, in
such models there would be a connection between the anomalies in
$\bsmumu$ and $\btopik$ decays, which is quite intriguing.

Many $Z'$ models have been proposed to explain the $\bsmumu$ anomalies
\cite{CCO, Crivellin:2015lwa, Isidori, dark_SSV, Chiang, Virto, GGH1,
  GGH2, BG, BFG, Perimeter, CDH1, CDH2, CHMNPR, CFJS, BDW, FNZ,
  Carmona:2015ena, AQSS, CFL, Hou, CFV, CFGI, IGG, Bhatia:2017tgo,
  BdecaysDM, ZMeV, Ko:2017lzd, Megias:2017ove, Ahmed:2017vsr,
  Kamenik:2017tnu, Sala:2017ihs, DiChiara:2017cjq, Alonso:2017bff,
  Bonilla:2017lsq, Alonso:2017uky, Chiang:2017hlj}. (Note that
Refs.~\cite{GGH1, GGH2, BG, BFG} all discuss the 3-3-1 model.  Here
the $Z'$ couples equally to $e^+e^-$ and $\mu^+\mu^-$, so this model
cannot explain $R_{K^{(*)}}$.) The question is: are there models in
which the $Z'$ couples to right-handed $d$ and/or $u$ quarks, with
$g_R^{dd} \ne g_R^{uu}$? A survey of the models reveals the following:
\begin{itemize}

\item Models in which the $Z'$ couplings to $u{\bar u}$ or $d{\bar d}$
  either vanish or are very small include Refs.~\cite{CCO, IGG,
    BdecaysDM, Alonso:2017uky, Chiang:2017hlj, Kamenik:2017tnu,
    Alonso:2017bff, Bonilla:2017lsq}.

\item Models in which the $Z'$ has vectorlike couplings to $u{\bar u}$
  and $d{\bar d}$ include Refs.~\cite{Crivellin:2015lwa, CDH2, FNZ,
    CFV, CFGI, Bhatia:2017tgo}.  In this case the $Z'$ model cannot
  explain the $\btopik$ puzzle, see Sec.~\ref{PNPnonzero}.

\item Models that focus only on $\bsmumu$ and say nothing about any
  other couplings include Refs.~\cite{CHMNPR, Ahmed:2017vsr,
    DiChiara:2017cjq}.

\item Models in which the $Z'$ has explicit couplings to RH quarks,
  with different couplings to RH $u{\bar u}$ and $d{\bar d}$ include
  the 3-3-1 model \cite{GGH1, GGH2, BG, BFG} and
  Refs.~\cite{Ko:2017lzd,CFJS}.

\item Many models have been proposed in which the $Z'$ couples to LH
  quarks, but not RH quarks. A significant fraction of these can be
  easily modified to allow the $Z'$ to couple to RH quarks. These
  include Refs.~\cite{Isidori, Chiang, Virto, dark_SSV, Perimeter,
    CDH1, BDW, Carmona:2015ena, AQSS, CFL, Hou, ZMeV, Megias:2017ove,
    Sala:2017ihs}.

\end{itemize}
There are a total of 34 $Z'$ models. Of these, 17 have, or can be
modified to have, the $Z'$ coupling to right-handed $u{\bar u}$ and/or
$d{\bar d}$, with $g_R^{dd} \ne g_R^{uu}$. These models can therefore
potentially also explain the $\btopik$ puzzle.

It is necessary to check other constraints. In order to reproduce the
$\bsmumu$ data, the $Z'$ mass and couplings must satisfy
\beq
C_9^{\mu \mu}({\rm NP}) = -C_{10}^{\mu \mu}({\rm NP}) = -\left[ \frac{\pi}{\sqrt 2 G_F \alpha V_{tb} V_{ts}^*} \right ] \,
\frac{g_L^{bs} g_L^{\mu \mu}}{M_{Z'}^2} ~.
\eeq
For $C_9^{\mu \mu}({\rm NP}) = -C_{10}^{\mu \mu}({\rm NP}) = -0.8$
\cite{Alok:2017jgr}, we require
\beq
\frac{|g_L^{bs} g_L^{\mu\mu}|}{M_{Z'}^2} = 1.2 \times 10^{-3} ~ {\rm TeV}^{-2} ~.
\label{constraint3}
\eeq
The $Z'$ will also contribute to $\nu_\mu N \to \nu_\mu N \mu^+ \mu^-$
(neutrino trident production). Measurements of this process lead to
the constraint \cite{Alok:2017jgr}
\beq
\frac{|g_L^{\mu\mu}|^2}{M_{Z'}^2} < 1.6 ~ {\rm TeV}^{-2} ~.
\label{constraint4}
\eeq

For example, in Ref.~\cite{Alok:2017sui}, the $\bsmumu$ data were
analyzed in the context of $Z'$ models. Good fits were found for
$M_{Z'} = 1$ TeV, with $g_L^{\mu\mu} \simeq 0.5$ and $g_L^{bs} \simeq
-2.5 \times 10^{-3}$. From Eq.~(\ref{NPcoupratio}), this implies that
the $Z'$ couplings to right-handed $u{\bar u}$ and/or $d{\bar d}$ must
be on the large side, $O(1)$.

However, there is another, more important process to consider. A $Z'$
that couples to both quarks and muons can be detected at the LHC via
$p p \to Z' \to \mu^+ \mu^-$. If the $Z'$ couples only to ${\bar s}b$,
its production cross section may be small enough to escape detection
(for example, see Ref.~\cite{BdecaysDM}). On the other hand, if the
$Z'$ couples to ${\bar d}d$ and/or ${\bar u}u$, it will be produced
plentifully in $pp$ collisions. Recently, using the 2015 and 2016 data
at $\sqrt{s} = 13$ TeV, the ATLAS Collaboration searched for high-mass
resonances decaying into dileptons, but found nothing
\cite{Aaboud:2017buh}. They put an upper limit on the product of the
production cross section and decay branching ratio, converting this to
a lower limit on $M_{Z'} \gsim 4$ TeV for the $Z'$ models analyzed. We
expect that this limit applies to our $Z'$.

The conclusion is that some of the $Z'$ models proposed to explain the
$\bsmumu$ data may also explain the $\btopik$ puzzle. The $Z'$ must be
somewhat massive, with $M_{Z'} \gsim 4$ TeV. On the other hand, its
mass cannot be {\it much} larger than this lower limit, as the
couplings would become nonperturbative in this case. So perhaps such a
$Z'$ will be observed at the LHC in the coming years.

\subsection{\bf Diquarks}

Another NP particle that can contribute to $\bsqq$ ($q=u,d$) at tree
level is a diquark \cite{GGS}. Diquarks are scalar particles that
couple to two quarks\footnote{In principle, diquarks can also be
  vector particles. In this case, it is natural to consider them to be
  the gauge bosons of an extended gauge group (such as SU(5)).
  However, in general such diquarks also have leptoquark couplings,
  which leads to proton decay. Scalar diquarks do not have this
  problem.}. Quarks are ${\bf 3}$s under $SU(3)_C$, so that the
diquark must transform as a ${\bf{\bar 3}}$ (triplet) or a ${\bf 6}$
(sextet). Since diquark couplings are fermion-number-violating, the
diquarks couple to two quarks of the same chirality. That is, they
couple to $q_L^i q_L^j$, $u_R^i u_R^j$, $u_R^i d_R^j$, or $d_R^i
d_R^j$, where $q_L = (u_L, d_L)$ is an $SU(2)_L$ doublet, and $i,j$
are flavor indices. There are therefore a total of 8 different types
of diquark. These are listed in Table \ref{diquarks}.

\begin{table}[tbh]
\center
\begin{tabular}{|c|c|c|c|c|}
\hline
Name & $SU(3)_C$ & $SU(2)_L$ & $U(1)_Y$ & $QQ$ Coupling \\ \hline
$I$ & ${\bf 6}$ & ${\bf 3}$ & $\frac23$ & $q_L^i q_L^j$ \\
$II$ & ${\bf{\bar 3}}$ & ${\bf 3}$ & $\frac23$ & $q_L^i q_L^j$ \\
$III$ & ${\bf 6}$ & ${\bf 1}$ & $\frac23$ & $q_L^i q_L^j$, $u_R^i d_R^j$ \\
$IV$ & ${\bf{\bar 3}}$ & ${\bf 1}$ & $\frac23$ & $q_L^i q_L^j$, $u_R^i d_R^j$ \\
$V$ & ${\bf 6}$ & ${\bf 1}$ & $\frac83$ & $u_R^i u_R^j$ \\
$VI$ & ${\bf{\bar 3}}$ & ${\bf 1}$ & $\frac83$ & $u_R^i u_R^j$ \\
$VII$ & ${\bf 6}$ & ${\bf 1}$ & $-\frac43$ & $d_R^i d_R^j$ \\
$VIII$ & ${\bf{\bar 3}}$ & ${\bf 1}$ & $-\frac43$ & $d_R^i d_R^j$ \\
\hline
\end{tabular}
\caption{Scalar diquarks: quantum numbers and couplings.}
\label{diquarks}
\end{table}

The triplets are antisymmetric in color, while sextets are
symmetric. This implies that their couplings are also respectively
antisymmetric and symmetric in flavor \cite{GGS}. This has important
consequences for flavor-changing neutral currents (FCNCs). Consider
the diquarks that couple to $d s$ (diquarks $I$, $II$, $VII$ and
$VIII$). In principle, they contribute at tree level to $K^0$-${\bar
  K}^0$ mixing via the t-channel exchange of a diquark. This involves
the transitions $d \to {\bar d}$ and ${\bar s} \to s$, i.e., the
flavor indices of the couplings have $i = j$. However, because the
couplings of the triplet diquarks are antisymmetric in flavor, these
couplings vanish for these diquarks. That is, while sextet diquarks
can generate $\Delta F=2$ FCNCs, triplet diquarks cannot \cite{GGS}.
Thus, the measurements of neutral meson mixing constrain diquarks $I$,
$V$ and $VII$ to be very massive, so that their effects in other
low-energy processes are negligible. However, there are no similar
constraints on diquarks $II$, $III$, $IV$, $VI$ and $VIII$.

Diquarks contribute at tree level to $\bsqq$ ($q=u,d$) via $b \to
{\bar q} D^* (\to s q)$. From this, we see that diquark $VI$ does not
contribute to $\btopik$. Recall that in Sec.~\ref{ModelindepNP}
we noted that there are four NP matrix elements that contribute to
$\btopik$ decays: $\ApNPqph$ and $\ApNPCqph$ ($q=u,d$). The
color-allowed and color-suppressed matrix elements use the operators
${\bar s}_\alpha \Gamma_i b_\alpha \, {\bar q}_\beta \Gamma_j q_\beta$
and ${\bar s}_\alpha \Gamma_i b_\beta \, {\bar q}_\beta \Gamma_j
q_\alpha$, respectively. Now consider diquark $VIII$. It contributes to
$\ApNPdph$ and $\ApNPCdph$. However, it is straightforward to see that
these are not independent. The $\bsqq$ transition involves $b_\alpha \to {\bar
  q}_\beta D^*_{\alpha\beta}$. The virtual $D^*$ then decays equally
to $s_\beta d_\alpha$ and $s_\alpha d_\beta$. But the first decay
creates the color-allowed operator, the second the color-suppressed
operator. And since the triplet is antisymmetric, there is a relative
minus sign. That is, we have $\ApNPCdph = -\ApNPdph$, or
$\Phi_d^{\prime {C}} = \Phi'_d$ and ${\cal A}^{\prime {C}, d} = -{\cal
  A}^{\prime,d}$. This same type of behavior holds for all diquarks,
except that diquark $III$ is a sextet, which is symmetric, so that
$\ApNPCuph = + \ApNPuph$.

The four diquarks that contribute to $\btopik$ are $II$, $III$, $IV$
and $VIII$. These have the following properties:
\begin{itemize}

\item $II$: decays to $q_L^i q_L^j$, which includes $u_L^i u_L^j$,
  $u_L^i d_L^j$ and $d_L^i d_L^j$. This implies that the amplitudes
  for $\bsuu$ and $\bsdd$ are equal, so that ${\cal A}^{\prime,d} =
  {\cal A}^{\prime,u}$ and ${\cal A}^{\prime {C}, d} = {\cal
    A}^{\prime {C}, u}$. We then have $\pewpnp = \pewcpnp = 0$; the
  only nonzero NP operator is $P'_{NP}$. (We also have $\ApNPCqph =
  -\ApNPqph$, but this is not important.)

\item $III$: has $\ApNPCdph = \ApNPdph = 0$ and $\ApNPCuph = +
  \ApNPuph$.

\item $IV$: has $\ApNPCdph = \ApNPdph = 0$ and $\ApNPCuph = -
  \ApNPuph$.

\item $VIII$: has $\ApNPCdph = -\ApNPdph$ and $\ApNPCuph = \ApNPuph =
  0$.

\end{itemize}
With this information, we can perform fits for the four diquark models.

\subsubsection{\bf\boldmath Diquark $II$}

The scenario where $P'_{NP}$ is the only nonzero NP operator was
examined in Sec.~\ref{PNPnonzero}. There it was found that the fit was
no better than that of the SM, and that the $\btopik$ puzzle could not
be explained.

\subsubsection{\bf\boldmath Diquark $III$}

The results of the fit with diquark $III$ are shown in Table
\ref{tab:NPfitDQIII}. The best fit has $\chi^2/{\rm d.o.f.} = 4.0/3$,
for a p-value of $26\%$. Like the $Z'$ model of Sec.~\ref{Z'cases324},
the fit is not bad (and is far better than the SM). It could explain the
$\btopik$ puzzle.

\begin{table}[tbh]
\center
\begin{tabular}{|c|c|}
\hline
\multicolumn{2}{|c|}{Diquark $III$: $\chi^2/{\rm d.o.f.} = 4.0/3$,}\\
\multicolumn{2}{|c|}{~~~~~~~~~~~~~~~~~~ p-value $=0.26$}\\
 \hline
Parameter & Best-fit value \\ 
\hline
$\gamma$ & $(72.1 \pm 5.8)^\circ$ \\
\hline
$\beta$ & $(21.87 \pm 0.68)^\circ$ \\
\hline
$\Phi'$ & $(273.2 \pm 10.5)^\circ$ \\
\hline
$|T'|$ & $7.8 \pm 5.5$ \\
\hline
$|P'_{tc}|$ & $50.6 \pm 0.7$ \\
\hline
$\pewpnp$ & $4.6 \pm 7.5$ \\
\hline
$\delta_{P'_{tc}}$ & $(317.8 \pm 47.5)^\circ$ \\
\hline
$\delta_{C'}$ & $(128.0 \pm 329.0)^\circ$ \\
\hline
\end{tabular}
\caption{$\chi^2_{\rm min}/{\rm d.o.f.}$ and best-fit values of
  unknown parameters for the model of diquark $III$. Constraints:
  $\btopik$ data, measurements of $\beta$ and $\gamma$, $|C'/T'| =
  0.2$.}
\label{tab:NPfitDQIII}
\end{table}

In order to get a sense of what values are required for the diquark
mass and couplings, we proceed as in Sec.~\ref{Z'cases324}: we compare
$\pewpnp$ and $T'$. Here, however, the form factors for the two
diagrams are different, so this comparison will only give a rough
idea. We have
\bea
&& \left\vert \frac{\pewpnp}{T'} \right\vert \approx
\frac{|g^{ub} g^{us}|/M_{DQ}^2}{(G_F/\sqrt{2})|V_{ub}^* V_{us}|} = \frac{4.6}{7.8} \nn\\
&\Longrightarrow& \frac{|g^{ub} g^{us}|}{M_{DQ}^2} = 4.5 \times 10^{-3} ~ {\rm TeV}^{-2} ~.
\eea
This is similar to what was found for $Z'$ models.

\subsubsection{\bf\boldmath Diquark $IV$}

The results of the fit with diquark $IV$ are shown in Table
\ref{tab:NPfitDQIV}. The best fit has $\chi^2/{\rm d.o.f.} = 5.4/3$,
for a p-value of $15\%$. This is only so-so. Given that the best-fit
values of the ratios $|T'/P'_{tc}|$ and $|\pewpnp/P'_{tc}|$ are both
somewhat larger than expected, we conclude that this diquark does not
provide a good explanation of the $\btopik$ puzzle.

\begin{table}[tbh]
\center
\begin{tabular}{|c|c|}
\hline
\multicolumn{2}{|c|}{Diquark $IV$: $\chi^2/{\rm d.o.f.} = 5.4/3$,}\\
\multicolumn{2}{|c|}{~~~~~~~~~~~~~~~~~~ p-value $=0.15$}\\
 \hline
Parameter & Best-fit value \\ 
\hline
$\gamma$ & $(69.9 \pm 5.6)^\circ$ \\
\hline
$\beta$ & $(21.86 \pm 0.68)^\circ$ \\
\hline
$\Phi'$ & $(0.6 \pm 6.8)^\circ$ \\
\hline
$|T'|$ & $31.5 \pm 4.8$ \\
\hline
$|P'_{tc}|$ & $50.3 \pm 0.8$ \\
\hline
$\pewpnp$ & $22.6 \pm 3.4$ \\
\hline
$\delta_{P'_{tc}}$ & $(186.3 \pm 0.9)^\circ$ \\
\hline
$\delta_{C'}$ & $(57.5 \pm 17.0)^\circ$ \\
\hline
\end{tabular}
\caption{$\chi^2_{\rm min}/{\rm d.o.f.}$ and best-fit values of
  unknown parameters for the model of diquark $IV$. Constraints:
  $\btopik$ data, measurements of $\beta$ and $\gamma$, $|C'/T'| =
  0.2$.}
\label{tab:NPfitDQIV}
\end{table}

\subsubsection{\bf\boldmath Diquark $VIII$}

The results of the fit with diquark $VIII$ are shown in Table
\ref{tab:NPfitDQVIII}. The best fit has $\chi^2/{\rm d.o.f.} =
11.6/3$, for a p-value of $0.9\%$. This is a very poor fit, worse than
that of the SM. This diquark cannot explain the $\btopik$ puzzle.

\begin{table}[tbh]
\center
\begin{tabular}{|c|c|}
\hline
\multicolumn{2}{|c|}{Diquark $VIII$: $\chi^2/{\rm d.o.f.} = 11.6/3$,}\\
\multicolumn{2}{|c|}{~~~~~~~~~~~~~~~~~~ p-value $=0.009$}\\
 \hline
Parameter & Best-fit value \\ 
\hline
$\gamma$ & $(64.8 \pm 6.5)^\circ$ \\
\hline
$\beta$ & $(21.80 \pm 0.68)^\circ$ \\
\hline
$\Phi'$ & $(-19.4 \pm 91.7)^\circ$ \\
\hline
$|T'|$ & $9.3 \pm 3.2$ \\
\hline
$|P'_{tc}|$ & $51.7 \pm 2.1$ \\
\hline
$\pewpnp$ & $1.3 \pm 2.1$ \\
\hline
$\delta_{P'_{tc}}$ & $(-11.4 \pm 4.4)^\circ$ \\
\hline
$\delta_{C'}$ & $(250.2 \pm 24.8)^\circ$ \\
\hline
\end{tabular}
\caption{$\chi^2_{\rm min}/{\rm d.o.f.}$ and best-fit values of
  unknown parameters for the model of diquark $VIII$. Constraints:
  $\btopik$ data, measurements of $\beta$ and $\gamma$, $|C'/T'| =
  0.2$.}
\label{tab:NPfitDQVIII}
\end{table}

\subsection{\bf Summary}

In the previous section, we saw that, if $|C'/T'| = 0.2$, the SM
cannot explain the $\btopik$ puzzle. The two types of NP that can
contribute to $\btopik$ at tree level are $Z'$ bosons and diquarks. In
this section, we found that either NP model can explain the puzzle if
$|C'/T'| = 0.2$. For $Z'$ models, the $Z'$ must couple to right-handed
$u{\bar u}$ and/or $d{\bar d}$, with $g_R^{dd} \ne g_R^{uu}$. Half of
the $Z'$ models proposed to explain the $\bsmumu$ anomalies have the
required $Z'$ couplings to $u{\bar u}$ and/or $d{\bar d}$.  As for
diquarks, the only one that works is a color ${\bf 6}$ that couples to
$ud$. For both NP models, the fits have p-values in the range 25-40\%.

\section{Conclusions}

There are four $\btopik$ decays -- $B^+ \to \pi^+ K^0$, $B^+ \to \pi^0
K^+$, $\bd \to \pi^- K^+$ and $\bd \to \pi^0 K^0$ -- whose amplitudes
obey a quadrilateral isospin relation. In the early 2000s, their
branching ratios and CP asymmetries (direct and indirect) were
measured, and it was noted that there was a tension between the
measurements and the SM. This was referred to as the ``$\btopik$
puzzle.'' Over the years, a number of analyses were done, attempting
to quantify the seriousness of the puzzle, and to identify the type of
new physics that can ameliorate the problem.

In the present paper, we perform an update of the $\btopik$ puzzle by
doing fits to the data using a diagrammatic decomposition of the
$\btopik$ amplitudes. We find that the key unknown parameter is
$|C'/T'|$, the ratio of color-suppressed and color-allowed tree
amplitudes. Theoretically, this ratio is predicted to be $0.15 \lsim
|C'/T'| \lsim 0.5$. If it is large, $|C'/T'| = 0.5$, we find that
the SM can explain the data: the fit has a p-value of 43\% (an
excellent fit has p-value $=50\%$). On the other hand, if it is small,
$|C'/T'| = 0.2$, the fit has a p-value of 4\%, which is poor. Our
conclusion is that, if $|C'/T'|$ is small, the SM cannot explain the
$\btopik$ puzzle -- NP is needed.

The two types of NP that can contribute to $\btopik$ at tree level are
$Z'$ bosons and diquarks. For the case of $|C'/T'| = 0.2$, we examine
whether the $\btopik$ puzzle can be explained with the inclusion of
such NP. For both types of NP, the answer is yes. 

In the case of $Z'$ models, the decay $\bsqq$ ($q=u,d$) is produced
via the tree-level exchange of a $Z'$ that couples to ${\bar s}b$ and
to ${\bar q}q$. We find that, if the $Z'$ couples only to left-handed
${\bar q}q$, things do not work. The $\btopik$ puzzle can be explained
only if the $Z'$ couples to right-handed $u{\bar u}$ and/or $d{\bar
  d}$, with $g_R^{dd} \ne g_R^{uu}$ -- the p-values of the fits are in
the range 25-40\%. 

This particular NP solution is intriguing because there are currently
anomalies involving the process $\bsmumu$ that can also be explained
by the addition of a $Z'$. We find that, of all the $Z'$ models
proposed for the $\bsmumu$ anomalies, half have the required $Z'$
couplings to $u{\bar u}$ and/or $d{\bar d}$. Such models could
potentially explain both the $\bsmumu$ anomalies and the $\btopik$
puzzle.

Turning to diquarks, there are eight different types. Taking into
account constraints from other processes, particularly $\Delta F=2$
FCNCs, there is only one diquark that works. It is a color ${\bf 6}$
that couples to $ud$. It contribute at tree level to $b \to s {\bar u}
u$ via $b \to {\bar u} D^* (\to s u)$. Its fit has a p-value of 26\%.

\bigskip
\noindent
{\bf Acknowledgments}: We thank A. Alok for help with MINUIT,
B. Bhattacharya for collaboration at the early stages of the project,
G. Bell for helpful discussions about the value of $|C'/T'|$ at NNLO,
and R. Watanabe for helpful discussions about diquarks.  This work was
financially supported by NSERC of Canada (NBB, DL, JR), and by the
National Science Foundation (AD, AR) under Grant No.\ NSF
PHY-1414345. AD acknowledges the hospitality of the Department of
Physics and Astronomy, University of Hawaii, where part of the work
was done.


\end{document}